\documentclass[10pt,conference]{IEEEtran}

\usepackage[many]{tcolorbox}
\usepackage{listings}
\usepackage{enumitem}
\usepackage{pifont}
\usepackage{amsfonts}
\usepackage{amsmath}
\usepackage{amsthm}
\usepackage{booktabs}
\usepackage{multirow}
\usepackage{makecell}
\usepackage{subcaption}
\usepackage{algorithm}
\usepackage{algpseudocode}
\usepackage{hyperref}
\usepackage{mfirstuc}
\usepackage[capitalise,nameinlink]{cleveref}
\usepackage{newtxmath}

\theoremstyle{definition}
\newtheorem{definition}{Definition}[section]

\newcommand{\code}[1]{\texttt{\small#1}}

\def\capitalizehelper#1#2{\ifx\relax#2\relax\else\uppercase{#1}\lowercase{#2}\fi}
\newcommand{\capitalize}[1]{\uppercase{\expandafter\capitalizehelper\string#1}}%

\newcommand{\tech}{RR-Reduce}
\newcommand{\hybrid}{Hybrid-Reduce}
\newcommand{\trait}{execution-aware}

\newcommand{\evalsize}{28}
\newcommand{\bugsize}{13}
\newcommand{\commanderkeensize}{1.3}

\gdef\wasmshrinkcode{12.88}
\gdef\wasmreducecode{0.43}
\gdef\rrreducecode{6.67}
\gdef\rrreducetarget{1.20}
\gdef\hybridcode{0.13}
\gdef\rrhybridbest{23}
\gdef\rrhybridtie{six}

\gdef\hybridimprovesize{3.42}

\gdef\rrreducetime{871}

\gdef\hybridimprovetime{2.26}
\gdef\hybridimprovetimeshrink{3.02}
\gdef\rrreduceimprovetime{33.15}

\begin{document}

\title{Execution-Aware Program Reduction for\\WebAssembly via Record and Replay}

\author{%
\begin{tabular}{ccc}
Doehyun Baek & Daniel Lehmann & Ben L. Titzer \\
doehyunbaek@gmail.com & mail@dlehmann.eu & btitzer@andrew.cmu.edu \\
University of Stuttgart & Google Germany GmbH & Carnegie Mellon University \\
Stuttgart, Germany & Munich, Germany & Pittsburgh, USA
\end{tabular}
\\[2ex]
\begin{tabular}{cc}
Sukyoung Ryu & Michael Pradel \\
sryu.cs@kaist.ac.kr & michael@binaervarianz.de \\
KAIST & CISPA Helmholtz Center \\
Daejeon, South Korea &  for Information Security \\
 & Stuttgart, Germany
\end{tabular}
}

\maketitle

\begin{abstract}
WebAssembly (Wasm) programs may trigger bugs in their engine implementations.
To aid debugging, program reduction techniques try to produce a smaller variant of the input program that still triggers the bug.
However, existing execution-unaware program reduction techniques struggle with large and complex Wasm programs, because they rely on static information and apply syntactic transformations, while ignoring the valuable information offered by the input program's execution behavior.
We present \tech{} and \hybrid{}, novel \trait{} program reduction techniques that leverage execution behaviors via record and replay.
\tech{} identifies a bug-triggering function as the target function, isolates that function from the rest of the program, and generates a reduced program that replays only the interactions between the target function and the rest of the program.
\hybrid{} combines a complementary execution-unaware reduction technique with \tech{} to further reduce program size.
We evaluate \tech{} and \hybrid{} on 28 Wasm programs that trigger a diverse set of bugs in three engines.
On average, \tech{} reduces the programs to \rrreducetarget{}\% of their original size in 14.5 minutes, which outperforms the state of the art by \rrreduceimprovetime{}$\times$ in terms of reduction time.
\hybrid{} reduces the programs to \hybridcode{}\% of their original size in 3.5 hours, which outperforms the state of the art by \hybridimprovesize{}$\times$ in terms of reduced program size and \hybridimprovetime{}$\times$ in terms of reduction time.
We envision \tech{} as the go-to tool for rapid, on-demand debugging in minutes, and \hybrid{} for scenarios where developers require the smallest possible programs.
\end{abstract}

\section{Introduction}
\label{sec:introduction}

WebAssembly (Wasm)~\cite{Haas2017} is a language for the web designed for efficient, sandboxed execution.
Wasm engines, like any other software, may contain bugs.
Indeed, several techniques for detecting bugs in Wasm implementations have been proposed recently~\cite{zhou2023wadiff,10.1145/3650212.3680358,zhao2024wapplique,rao2024diffspec}.
These bug-triggering Wasm programs are often large and complex, which makes debugging the affected engine challenging.

Program reduction, which aims to find a smaller variant of the input program that still triggers the bug, is used to mitigate this problem.
Delta Debugging~\cite{DBLP:journals/tse/ZellerH02} is the pioneering input reduction technique, and various others have been proposed since then~\cite{DBLP:conf/icse/Misherghi06,Herfert2017AutomaticallyRT,Sun2018PersesSP}.
For reducing Wasm programs, there are two industrially supported tools: wasm-reduce~\cite{wasm-reduce} and wasm-shrink~\cite{wasm-shrink}.
While these tools are effective in some cases, they are limited in terms of both effectiveness and efficiency.

To illustrate the limitations of currently available techniques, consider the \emph{commanderkeen} program~\cite{wasm-r3}, which reveals a miscompilation bug in the Wizard engine~\cite{wizard}.
The Wasm binary is 3.9MB in size, which is not uncommon for binaries compiled from large programs.
Wasm-reduce, i.e., the most effective existing tool for Wasm, reduces commanderkeen to \commanderkeensize{}MB after 24 hours of processing, which is far too large for a human to manually debug.
Wasm-reduce is an example of an execution-unaware program reduction technique, i.e., it ignores the execution behavior of the input program.
We identify two reasons why such execution-unaware techniques struggle with our example program and other input programs:
(1)~The reduction process applies syntactic transformations based on static information only, leading to a huge search space of possible reduced programs.
(2)~Checking whether the bug is preserved by a reduced program is expensive, as the Wasm engine has to execute the program to check whether the bug is still triggered.
For the commanderkeen program, each such oracle invocation takes about 9 seconds, which is a long time for a program reduction technique that needs to perform many oracle invocations to find a reduced program.

A detailed inspection of the above example and other bug-triggering Wasm programs reveals an interesting observation:
Usually, a single function in the program is responsible for triggering the bug.
For example, the commanderkeen program contains 1,970 functions, but only one of them is relevant for triggering the bug.
A naive approach to reduce the program would be to simply remove all other functions from the Wasm binary.
However, this naive approach would lead to a program that fails to compile and execute, hence failing to trigger the bug, as the bug-triggering function depends on other functions and other code in the program.
For the motivating example, 580 other functions are executed before the bug is triggered.

We hypothesize that these problems can be addressed by leveraging the execution behavior of the input program, which provides valuable information about the dependencies between different parts of the program.
This information may be used for isolating the bug-triggering function and any other code necessary to trigger a bug-exposing execution.
Based on this hypothesis, we ask the question: Can we use information about the execution of an input program to reduce the input program more effectively?

To answer this question, this paper presents \tech{}, an \emph{\trait{}} program reduction technique for Wasm.
A key insight of the approach is that we can repurpose existing record and replay techniques for \trait{}~program reduction.
Record and replay is a debugging technique that records a program execution and then replays the recorded execution.
It has been adopted in many domains, including native binaries~\cite{PinPlay,rr}, JavaScript~\cite{jsbench,jalangi}, and Wasm~\cite{wasm-r3}.
Usually, record and replay is used to provide a deterministic way to replay an entire execution of a program.
Instead, we use this technique to replay only parts of an input program that are essential to trigger the bug.

More specifically, \tech{} performs three steps:
First, identify a bug-triggering function of the input program, which we call the target function.
Second, partition the input program into two subprograms: one containing the target function and another containing the remaining functions.
Third, use record and replay to generate a new program that includes the unmodified target function and generated replay functions that mimic the behavior of the remaining functions.
Our \tech{} approach can be used as a stand-alone program reduction technique, or in combination with existing program reduction techniques.
We call the latter approach \hybrid{}, which applies an existing reduction technique (wasm-reduce) to the output of \tech{}.
This further reduces output, at the expense of additional time.

\begin{figure}
      \includegraphics[width=\linewidth]{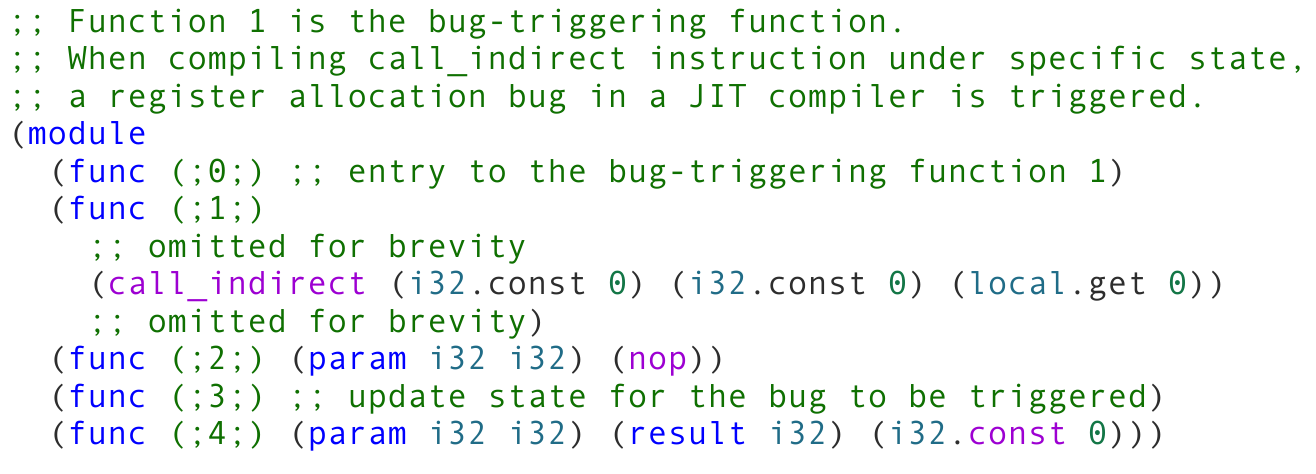}
      \caption{\hybrid{}'s output for commanderkeen.}
      \label{f:commanderkeen-hybrid}
\end{figure}

Getting back to our motivating example, Figure~\ref{f:commanderkeen-hybrid} shows the reduced program produced by \hybrid{}.
Instead of the \commanderkeensize{}MB output of the existing wasm-reduce tool, our approach yields a reduced program of only 158 bytes.
Such a small program enables engine developers to debug the bug in a reasonable time, ultimately leading to more robust implementations of Wasm.

We evaluate \tech{} and \hybrid{} on 28 Wasm programs that trigger bugs in three Wasm engines.
The results show that \hybrid{} clearly outperforms the state of the art, wasm-reduce, and that our two approaches offer different effectiveness-efficiency trade-offs.
Specifically, \tech{} reduces the amount of code a developer must inspect to \rrreducetarget{}\% of the original size, on average, while taking \rrreducetime{} seconds, which is a \rrreduceimprovetime{}x improvement in efficiency over the state of the art.
That is, \tech{} significantly reduces input programs while offering unprecedented efficiency.
The \hybrid{} approach further reduces the program to \hybridcode{}\% of its original size, while taking about 3.5 hours, on average.
With these results, \hybrid{} outperforms the state of the art both in terms of the resulting program size (\hybridimprovesize{}$\times$ improvement) and in terms of reduction time (\hybridimprovetime{}$\times$ improvement).

In summary, this paper contributes the following:
\begin{itemize}
\item The novel idea of leveraging targeted record and replay for execution-aware program reduction.
\item A concrete implementation of this idea in \tech{} and \hybrid{}, which are two program reduction techniques for Wasm.
\item Empirical evidence that the use of execution behavior leads to more effective and efficient outcomes compared to existing program reduction techniques.
\item We release our tools and data as open source, for others to reproduce and build on our results: https://github.com/sola-st/rr-reduce.
\end{itemize}

\section{Background}
\label{s:backgrounds}

\subsection{WebAssembly}
\label{s:webassembly}

\newcommand\X[1]{\mathit{#1}}
\newcommand\K[1]{\textsf{\textbf{#1}}}

\begin{figure}[t]
\def\baselinestretch{1.2}
\footnotesize
$$
\begin{array}{@{}rrlrrl@{}}
\X{module} &::=& \X{function}^\ast~\X{global}^\ast~\X{table}^\ast~\X{memory}^\ast & \\
\X{function} &::=& \X{type_{func}}~(\X{import}~|~\X{code})~\X{export}^\ast & \\
\X{global} &::=& \X{type_{val}}~(\X{import}~|~\X{init})~\X{export}^\ast & \\
\X{table} &::=& \X{import}^?~\X{idx_{func}}^\ast~\X{export}^\ast & \\
\X{memory} &::=& \X{import}^?~\code{\footnotesize byte}^\ast~\X{export}^\ast & \\
\X{code} &::=& (\K{local}~\X{type_{val}})^\ast~\X{instr}^\ast & \\
\X{init} &::=& \X{instr}^\ast \\
\X{instr} &::=& \lefteqn{ \X{type_{val}}\K{.const}~\X{value}~\mid~ \X{type_{val}}\K{.load}~} \\
&\mid& \lefteqn{~\X{type_{val}}\K{.store}~\mid~\K{call}~\X{idx_{func}}~\mid} & \\
&\mid& \lefteqn{~\K{call\_indirect}~\X{type_{func}}~\mid~\K{return}~\mid~\cdots} & \\
\X{import} &::=& (\code{\footnotesize "module"}, \code{\footnotesize "name"}) \\
\X{export} &::=& (\code{\footnotesize "name"}) \\
\X{type_{func}} &::=& \X{type_{val}}^\ast \rightarrow \X{type_{val}}^\ast \\
\X{type_{val}} &::=& \K{i32}~\mid~\K{i64}~\mid~\K{f32}~\mid~\K{f64} \\
\X{idx_{func}} &\in & \mathbb{N} \\
\end{array}
$$
\caption{Abstract syntax of a simplified form of Wasm~\cite{Wasabi}.}\label{t:mini-wasm}
\end{figure}

Wasm~\cite{Haas2017} is a compact binary format designed for efficient sandboxed execution in modern web browsers.
Figure~\ref{t:mini-wasm} shows a simplified abstract syntax of Wasm.
A \emph{module}, representing a single binary file, contains functions, global variables, tables, and memories.
A \emph{function} accepts parameters, declares local variables, executes instructions, and returns results.
A \emph{global} variable stores a single value accessible across all functions and can be mutable or immutable.
A \emph{table} maps indices to references of host objects or Wasm functions.
A \emph{memory} is a contiguous, byte-addressable, page-sized mutable array.
These components can be imported from a host environment using module and name pairs, or exported under one or more names for external access.
Additionally, a module may include initialization data for tables and memories.

\subsection{Wasm-R3}
\label{s:wasm-r3}

Wasm-R3~\cite{wasm-r3} is a record and replay technique for Wasm.
It proceeds in three phases: record, reduce, and replay.
The record phase captures the execution of an input Wasm program and produces a trace of events.
The reduce phase minimizes this trace to a smaller sequence of events necessary for replay.
The replay phase generates replay functions that reproduce the behavior of host functions using the reduced trace, and merges them with the input Wasm program to create a replay program.
Of particular relevance to \tech{} is the replay phase.
The replay program generated by Wasm-R3 has the following characteristics:
(1) It preserves the individual functions in the input Wasm program as-is and only adds replay functions alongside them.
This differs from other record and replay techniques~\cite{jalangi} that modify the input program's functions, necessitated by the use of instrumentation during replay.
(2) The replay functions are implemented as standard Wasm code.
This approach makes the entire replay program standalone, allowing it to run on any Wasm engine and ensuring compatibility with various Wasm tools.
These characteristics enable us to apply an existing program reduction technique (wasm-reduce) to the output of Wasm-R3, which would not be feasible with record and replay techniques lacking these properties.

\section{Approach}
\label{s:approach}

This section presents \tech{}, an \trait{}~program reduction technique for Wasm via record and replay.
The approach leverages the ability of a record and replay technique to create replay functions that accurately reproduce an execution.
The key idea is not to replay the entire program, but to selectively replay those parts of the program that are relevant to triggering the bug.
We first give a high-level problem statement (Section~\ref{s:problem-statement}), provide an overview of our approach (Section~\ref{s:overview}), introduce the running example (Section~\ref{s:running}), and finally present the details of each step  (Sections~\ref{s:heuristics} to~\ref{s:validation}).

\subsection{Problem Statement}
\label{s:problem-statement}

We start by defining the problem we are addressing and by comparing it to problems addressed by previous work.
The input to any program reduction technique is a program:

\begin{definition}[Program]
  A program $p \in P$ is a sequence of functions $f_1, ..., f_m$, where each function $f_i$ consists of a sequence of instructions $i_1, ..., i_n$.
    The size of a program is the sum of the bytes of each instruction, i.e., $\mathit{size}(p) = \sum_{f_i \in p} \sum_{i_j \in f_i} \mathit{bytes}(i_j)$.
\end{definition}

In practice, programs may contain other information (Figure~\ref{t:mini-wasm}), such as initial values of global variables, which we ignore for the purpose of concisely defining the problem.
Our approach handles all elements of programs in Wasm, including functions, global variables, tables, and memories.
We also assume that the program is self-contained and does not require any additional inputs for execution.

The motivation for reducing a program is that it has some property of interest, such as triggering a bug in a runtime engine.
We assume to have an oracle that can check whether a program has this property.
For example, such an oracle can be implemented by running the program and checking whether it triggers a specific bug in the runtime engine.

\begin{definition}[Oracle]
An oracle $o: P \rightarrow \{ \mathit{true}, \mathit{false} \}$ is a function that yields $\mathit{true}$ if $p$ has the property of interest, and $\mathit{false}$ otherwise.
\end{definition}

Previous program reduction techniques, such as \cite{DBLP:journals/tse/ZellerH02,DBLP:conf/icse/Misherghi06,Herfert2017AutomaticallyRT,Sun2018PersesSP,Regehr2012TestcaseRF}, aim to reduce the size of a program while preserving the property of interest.
This problem can be formulated as follows:

\begin{definition}[Program reduction]
  Given a program $p \in P$ and an oracle $o$, where $o(p) = \mathit{true}$, find a reduced program $p'$ so that $o(p') = \mathit{true}$ and $\mathit{size}(p) > \mathit{size}(p')$.
\end{definition}

In this work, we consider not only the program itself but also an execution of the program.
Such an execution is available in many usage scenarios of program reduction, and it may provide valuable information for reducing a program.
For example, when debugging a bug in a runtime engine that is triggered by a specific input program, considering the execution of the input program that leads to the bug may be helpful for finding a smaller input program.
Our work exploits this insight by addressing a new, \trait{}~variant of the program reduction problem:

\begin{definition}[\xcapitalisewords{\trait{}} program reduction]
  Given a program $p \in P$, an execution $e$ of $p$ that triggers the property of interest (typically a bug), and an oracle $o$, where $o(p) = \mathit{true}$, find a reduced program $p'$ so that $o(p') = \mathit{true}$ and $\mathit{size}(p) > \mathit{size}(p')$.
\end{definition}

\begin{figure*}[htbp]
  \centering
  \includegraphics[width=\textwidth]{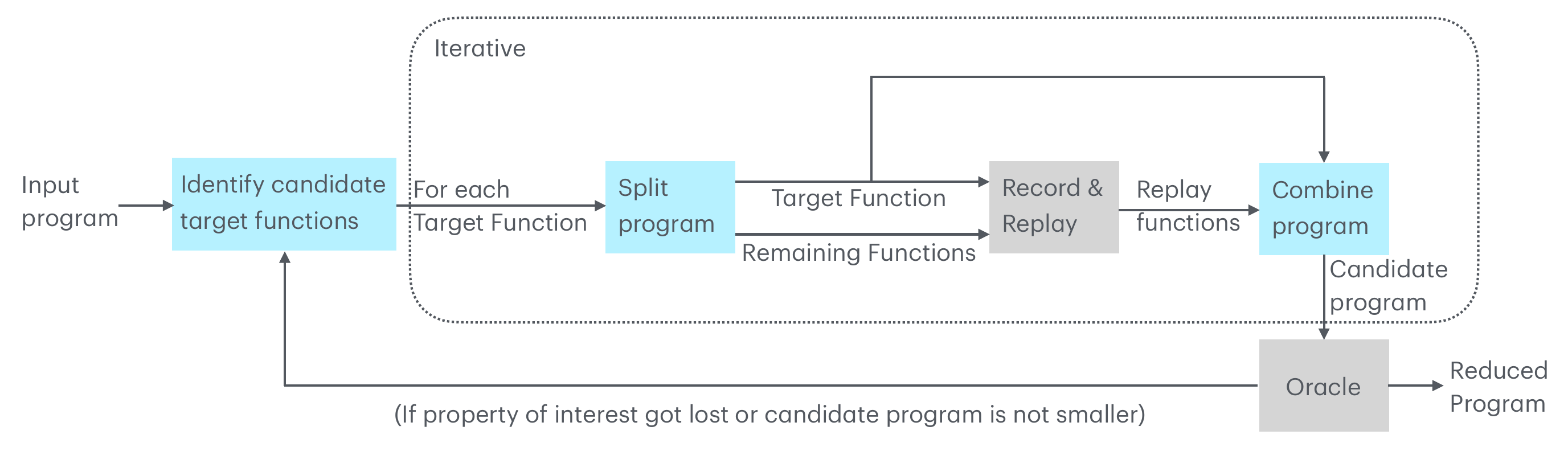}
  \caption{Overview of \tech{}. Components in blue are introduced in this paper, while components in gray are external.
  }
  \label{fig:overview}
\end{figure*}


To the best of our knowledge, we are the first to formulate and address the \trait{}~program reduction problem in this way.
The key difference between the \trait{}~program reduction problem and the standard program reduction problem is that the former considers an execution of the input program, while the latter does not.

In addition to the problem statement above, a practical program reduction technique should also fulfill two additional requirements:
\begin{enumerate}
  \item The reduced program should be small.
  Prior work on general test input reduction often strives for 1-minimality~\cite{DBLP:journals/tse/ZellerH02}, i.e., no single constituent of the input can be removed without losing the property of interest.
  Concretely, this could, e.g., mean that a reduction technique removes as many functions from a program as possible, while still preserving the property.
  Our evaluation will show that considering the execution behavior during program reduction can lead to smaller reduced programs than existing techniques~(Section~\ref{s:rq1}).
  \item The reduction process should be reasonably fast.
  Reasoning about an execution of a program can be computationally expensive, yet offers the potential to reduce programs faster due to the additional information available.
  Our evaluation will show that an \trait{}~reduction technique is not only more effective, but also more efficient than prior work~(Section~\ref{s:rq2}).
\end{enumerate}

\subsection{Overview}
\label{s:overview}

Figure~\ref{fig:overview} shows an overview of \tech{}.
Given an executable program as input, \tech{} first heuristically identifies functions in the program that may be critical for triggering the bug, which the approach considers as candidate target functions (Section~\ref{s:heuristics}).
Then, \tech{} iterates over the candidate target functions one by one, selecting each as the target function.
To start an iteration, the approach splits the input program into two parts: one containing the \emph{target function} and another containing all \emph{remaining functions} (Section~\ref{s:split}).
The resulting partitioned program is then passed to an existing record and replay technique, which records the execution and produces \emph{replay functions} (Section~\ref{s:rr}).
Importantly, we configure the record and replay technique to focus the replay on replaying only the target function.
That is, the remaining functions are either removed or replaced with replay functions that mimic those parts of the behavior of the original functions that are necessary to accurately replay the target function.
Given the replay functions, \tech{} combines the target function with the replay functions to produce a candidate for the reduced program (Section~\ref{s:combine}).
Finally, the oracle, also given as input, checks whether the candidate program still triggers the bug (Section~\ref{s:validation}).
If so, and if the candidate program is smaller than the input program, \tech{} returns the candidate program as the reduced program.
Otherwise, the approach repeats the iteration with a different target function.

\subsection{Running Example}
\label{s:running}

\begin{figure*}[htbp]
  \centering
  \includegraphics[width=1.03\linewidth]{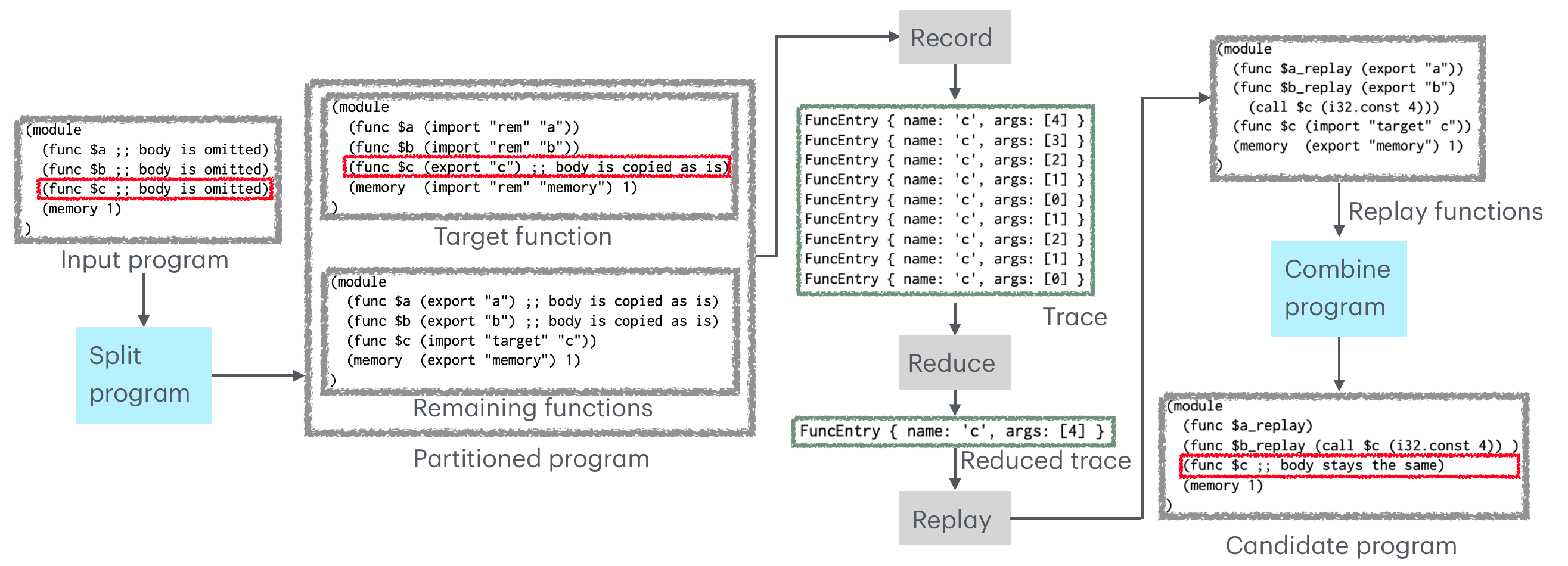}
  \caption{Running example of \tech{} for a single iteration.
  We assume the target function is already selected by the previous step.
  Red rectangles denote the target function.
}
\label{f:running-example}
\end{figure*}

Figure~\ref{f:running-example} shows a running example of \tech{} for a single iteration.
On the left, we have an input program that consists of three functions: \code{a}, \code{b}, and \code{c}.
Suppose an execution of this input program triggers a wrong-code bug in a Wasm engine, which results in a stack trace that contains the \code{c} function.
Because \code{c} is in the stack trace, \tech{} heuristically identifies \code{c} as the target function.
Next, \tech{} splits the input program into two parts: one containing the target function \code{c} and another containing the remaining functions \code{a} and \code{b}.
Then, \tech{} uses an existing record and replay technique, Wasm-R3 (Section~\ref{s:wasm-r3}), to produce replay functions that accurately replay the execution of the target function \code{c}.
This is achieved in three phases: record, reduce, and replay.
The resulting replay functions mimic the behavior of the remaining functions \code{a} and \code{b} in the input program.
Our approach combines these replay functions with the target function \code{c} to produce a candidate program, and checks using the oracle whether the candidate program still triggers the bug in the Wasm engine and is smaller than the input program.

\subsection{Identifying Candidate Target Functions}
\label{s:heuristics}


The first step of the approach is to identify a target function that triggers the bug.
Various heuristics could be used for this purpose.
For compiler-crash bugs in Wasm engines, our approach identifies the target function using error messages emitted by the compiler that specify which functions were being compiled at the time of the crash.
For wrong-code bugs, which typically manifest as runtime traps instead of the expected program execution, \tech{} identifies relevant functions through the program's stack trace captured when traps are triggered.
This heuristic stems from the observation that functions appearing in stack traces are often critical for both reproducing and diagnosing wrong-code bugs~\cite{DBLP:conf/issta/BurgerZ11,Moreno2014,Gu2019}.

To realize this heuristic, \tech{} maintains three sets of functions, where each function is identified by its function index:
The \emph{All} set, with all functions in the input program.
The \emph{Dynamic} set, with all functions that are executed in the given execution, which we obtain via a simple dynamic analysis.
The \emph{Heuristic} set, with functions deemed most likely to be relevant for reproducing an engine bug.
To compute the heuristic set, the approach performs a string search over the output of the engine compilation and execution, searching for function indices.
Given the three sets of functions, \tech{} enumerates candidates for the target function by prioritizing the \emph{Heuristic} set, then the \emph{Dynamic} set, and finally the \emph{All} set.

\subsection{Splitting the Program Into Target Function and Remaining Functions}
\label{s:split}

Next, \tech{} takes an input program and the index of the candidate target function as an input, and outputs a partitioned program, which consists of two newly created Wasm binaries: one containing the target function and the other containing the remaining functions.
To do so, the approach iterates over the program elements in the Wasm binary of the input program, performing different actions based on the type of each program element.
For functions, \tech{} copies each function into one of the two new binaries:
For the target function, it is copied to the binary that contains the target function.
For the remaining functions, they are copied to the binary that contains the remaining functions.
In addition to this copying, \tech{} performs additional bookkeeping to adjust the exports and imports of the newly created binaries.
Previously, all the functions were inside a single program, so no exports and imports were needed for them to call each other.
In contrast, in the partitioned program, each function is exported by one binary and then imported by the respective other binary.
For globals, tables, and memories, \tech{} always copies them to the binary containing the remaining functions, and then shares them between the two binaries via exporting and importing.

To see how this looks in practice, refer to our running example in Figure~\ref{f:running-example}, which shows how the input program gets split into two parts.
The input program on the left consists of three functions: \code{a}, \code{b}, and \code{c}.
On its right, the partitioned program consists of two binaries: The binary shown at the top, which contains the target function, and the binary shown at the bottom, which contains all the remaining functions of the input program.
Suppose we run \tech{} on this program with the target function set as \code{c}.
This means that function \code{a} and function \code{b} belong to the remaining functions.
We now iterate over program elements in the input program, starting from the functions.
For the \code{a} and \code{b} functions, as they belong to the remaining functions, we copy them to the binary at the bottom.
We also export \code{a} with name ``a'' and \code{b} with name ``b''.
In the binary at the top, we add imports with module ``rem'', name ``a'' and ``b'' respectively.
Next, for the \code{c} function, as it is the target function, we copy it to the binary at the top.
We also export \code{c} with name ``c''.
In the binary at the bottom, we add an import with module ``target``, name ``c''.
Finally, we move on to the memory section.
As the two binaries should share the same memory, we copy the memory to the binary at the bottom.
We also export memory with name ``memory''.
In the binary at the top, we add an import with module ``rem'', name ``memory''.

\subsection{Record and Replay for Program Reduction}
\label{s:rr}

Next, \tech{} applies the existing record and replay technique Wasm-R3 to the partitioned program.
This step generates replay functions that mimic the relevant behavior of the remaining functions.
A key difference between \tech{} and standard record and replay techniques is that \tech{} generates replay functions that replace functions belonging to the input program, instead of only reproducing side effects of the host environment.
This unlocks the use of record and replay for program reduction, as removing and replacing functions of the input program potentially reduces the program size.

Although Wasm-R3~(Section~\ref{s:wasm-r3}) itself is not a contribution of this paper, to better convey how \tech{} works, we illustrate how Wasm-R3 works for our running example in Figure~\ref{f:running-example}.
As a result of the split, we have a partitioned program with its two constituent parts.
Assume that \code{c} is the target function, and \code{a} and \code{b} are the remaining functions; \code{a} calls \code{b}, and \code{b} calls \code{c}.
Suppose additionally that the \code{c} function contains recursive calls to itself.
The record-reduce-replay part of the figure shows the three phases of Wasm-R3 applied to the partitioned program.
When Wasm-R3 is run with this partitioned program, it first records the target function's execution into a trace.
In the execution, the \code{c} function is called with an argument of 4 from the \code{b} function.
This leads to recursive calls of the \code{c} function, resulting in 9 function entry events in the trace.
However, as not all of these events are needed to replay the original execution of the target function, Wasm-R3 reduces the events to one, resulting in just a single function entry event with an argument of 4 in the reduce phase.
Lastly, in the replay phase, Wasm-R3 transforms this reduced trace into a replay function, which calls the \code{c} function with the argument 4.

\subsection{Combining Target Function and Replay Functions}
\label{s:combine}

After generating the replay functions, \tech{} combines the replay functions and the target function to produce a candidate program.
This is possible as the replay functions are identical to the remaining functions they replace in terms of imports and exports.
\tech{} then statically links and combines the replay functions and the target function into a single candidate program.
This process typically results in a reduced program as:
(1) Most of the remaining functions do not directly interact with the target function and are therefore eliminated.
(2) For the remaining functions that do interact with the target function, their replay counterparts are generally smaller, as they only need to replicate the behavior relevant to the target function, not their entire behavior.

To illustrate this step, consider our running example in Figure~\ref{f:running-example} again.
Replay functions do not contain their original bodies anymore; they contain the replay of the interactions with the target function.
Despite these changes, they are still exported with the same names, allowing \tech{} to statically link and combine them.
The resulting candidate program is presented on the right.
Comparing the candidate program with the input program, we see why \tech{} can act as an effective program reduction technique.
The target function is preserved, while one of the remaining functions has an empty body now, and the other remaining function has only two instructions.

\subsection{Validation of the Candidate Program}
\label{s:validation}

Finally, \tech{} validates the candidate program using the oracle.
Oracles are user-supplied scripts that check whether a program triggers the bug.
Although the oracle can be any script, in our evaluation, we use a simple Python script that runs the candidate program once in an engine with a bug and once in an engine without the bug.
Then, the oracle checks the return code, stdout, and stderr of the two runs to determine whether the candidate program still triggers the bug in the engine with the bug and terminates normally in the other engine.
If the candidate program still triggers the bug, and if the candidate program is smaller than the input program, \tech{} returns the candidate program as the reduced program.
Otherwise, the approach repeats the process with a different target function.
If no reduced program is found after trying all possible target functions, \tech{} returns the input program as the result.

\subsection{Combination with Existing Techniques into \hybrid{}}
\label{s:imp-hybrid}

The approach, as described so far, can be used as a stand-alone program reduction technique, which we call \tech{}.
In addition, \tech{} can be combined with existing program reduction techniques to potentially produce even smaller reduced programs.
We realize such a combination in \hybrid{}, which feeds the output of \tech{} into the existing wasm-reduce~\cite{wasm-reduce}.
Choosing between \tech{} and \hybrid{} is a trade-off between obtaining a reduced program that keeps the target function as-is and a potentially even smaller reduced program.
We study this trade-off in detail in our evaluation, which compares both approaches against the state of the art.


\section{Evaluation}
\label{s:evaluation}
We evaluate \tech{} by addressing the following three research questions:

\begin{itemize}[labelindent=\parindent,leftmargin=*]
\item \textbf{RQ1: Effectiveness}.
How effective are \tech{} and \hybrid{} in reducing input programs?
\item \textbf{RQ2: Efficiency}.
How efficient are \tech{} and \hybrid{} in terms of the time they take?
\item \textbf{RQ3: Qualitative Analysis}.
How do existing approaches and \tech{} qualitatively differ in their approach toward program reduction?
\end{itemize}

\subsection{Experimental Setup}
\label{s:setup}

\paragraph{Dataset}

As there is no previously available benchmark to evaluate Wasm program reduction techniques, we collect \evalsize{} Wasm programs that reveal \bugsize{} unique bugs in three Wasm engines: Wizard~\cite{WizardEngine}, WasmEdge~\cite{wasmedge}, and WAMR~\cite{Wamr}.
To construct the benchmark, we first gather 21 out of 27 Wasm programs from the Wasm-R3-Bench dataset~\cite{wasm-r3}, which are able to trigger nine bugs in the Wizard engine, and all 25 Wasm programs from the WASMaker paper~\cite{10.1145/3650212.3680358}, which trigger four bugs in WAMR and WasmEdge.
Out of the 21+25=46 candidate programs, we exclude 18 programs: 17 that use the SIMD extension of Wasm, which is currently not supported by Wasm-R3, and one that contains only a single function, i.e., there is nothing reduce for our approach.
Table~\ref{t:eval_set} lists the resulting \evalsize{} Wasm programs.
In addition to the programs themselves, we provide for each program an oracle script that checks whether a reduced program still triggers the same bug.

\paragraph{Metrics}
We measure the size of a Wasm program in terms of its code size, which is the size of the code section of the Wasm program.
This excludes the size of the data section, which contains initialization data for the memory, and the size of the custom section, which contains debug symbols.
Focusing on the code size is motivated by the fact that the code section is what engine developers typically focus on when debugging engine bugs.
For \tech{}, we report two kinds of code sizes: the ``All'' code size, which includes all functions in the binary, and the ``Target'' code size, which is the size of the target function.
The ``Target'' is more relevant for engine developers, as they only need to inspect the target function which is responsible for triggering the bug.

\setlength{\tabcolsep}{2pt}
\begin{table}[t]
  \centering
  \small
  \caption{\label{t:eval_set}Benchmark used to evaluate \tech{}.
  “Engine crash” refers to cases where the engine raises an error before executing the Wasm module.
  “Wrong code” refers to cases where the engine loads and compiles the Wasm module without errors but then deviates from the expected execution.
  }
  \begin{tabular}{llllr}
      \toprule
      Name             & Faulty engine & Fixed by & Kind & Code size \\
      \midrule
      wasmedge\#3018   & WasmEdge      & 93fd4ae  & Wrong code &   1,913 \\
      wamr\#2789       & WAMR          & 718f067  & Engine crash           &  17,604 \\
      wasmedge\#3019   & WasmEdge      & 93fd4ae  & Wrong code &  19,098 \\
      wamr\#2862       & WAMR          & 0ee5ffc  & Wrong code &  19,727 \\
      wamr\#2450       & WAMR          & e360b7a  & Engine crash           &  24,482 \\
      wasmedge\#3076   & WasmEdge      & 93fd4ae  & Wrong code &  31,365 \\
      mandelbrot       & Wizard        & 0b43b85  & Wrong code &  64,515 \\
      pathfinding      & Wizard        & ccf0c56  & Wrong code & 180,026 \\
      pacalc           & Wizard        & 81555ab  & Wrong code & 238,902 \\
      wasmedge\#3057   & WasmEdge      & 93fd4ae  & Wrong code & 243,564 \\
      guiicons         & Wizard        & 6d2b057  & Wrong code & 285,840 \\
      rtexviewer       & Wizard        & 708ea77  & Engine crash           & 296,617 \\
      rfxgen           & Wizard        & 6d2b057  & Wrong code & 378,918 \\
      riconpacker      & Wizard        & 6d2b057  & Wrong code & 398,627 \\
      rguistyler       & Wizard        & 6d2b057  & Wrong code & 410,845 \\
      rguilayout       & Wizard        & 6d2b057  & Wrong code & 416,692 \\
      jqkungfu         & Wizard        & 4e3e221  & Engine crash           & 487,607 \\
      bullet           & Wizard        & f7aca00  & Engine crash           & 536,115 \\
      funky-kart       & Wizard        & 6d2b057  & Wrong code & 607,293 \\
      sqlgui           & Wizard        & 6d2b057  & Wrong code & 628,046 \\
      hydro            & Wizard        & 708ea77  & Engine crash           & 719,538 \\
      figma-startpage  & Wizard        & 33ec201  & Engine crash           & 882,961 \\
      sandspiel        & Wizard        & ccf0c56  & Wrong code & 919,085 \\
      parquet          & Wizard        & 33ec201  & Engine crash           & 1,731,592 \\
      commanderkeen    & Wizard        & bc135ad  & Wrong code & 3,914,616 \\
      jsc              & Wizard        & 6d2b057  & Wrong code & 4,342,199 \\
      boa              & Wizard        & 6d2b057  & Wrong code & 5,198,069 \\
      ffmpeg           & Wizard        & 4e3e221  & Engine crash           & 5,356,751 \\
      \bottomrule
\end{tabular}
\end{table}

\paragraph{Baselines}

We compare \tech{} against two baselines: wasm-reduce~\cite{wasm-reduce} version 117, and wasm-shrink~\cite{wasm-shrink} version 1.227.0.
To our knowledge, wasm-reduce and wasm-shrink are the only existing program reduction techniques for Wasm, and hence, the current state of the art.
Both baselines are designed specifically for Wasm, following the style of C-Reduce~\cite{Regehr2012TestcaseRF}.
Wasm-reduce is part of the Binaryen toolchain~\cite{binaryen}.
It interleaves semantics-destroying reductions, such as replacing a node with its child, and semantics-preserving optimizations.
%
Wasm-shrink is part of the wasm-tools toolchain~\cite{wasm-tools}.
It also interleaves destructive reductions, such as deleting entire function bodies, with optimization reductions.
Like our approach, wasm-reduce and wasm-shrink take as an input a program to reduce and an oracle script.
Unlike \tech{}, neither of the two baselines are execution-aware, i.e., they do not make use of execution behavior of input programs.

\paragraph{Implementation and Hardware}
We implement \tech{} through a combination of Python, Rust, and JavaScript, building on several libraries and tools in the Wasm ecosystem.
The first step is to identify candidates for the target function (Section~\ref{s:heuristics}).
We use the existing tool, wasm-tools objdump~\cite{wasm-tools}, to compute the \emph{All} set,
a dynamic analysis based on the Wizard engine~\cite{wizard} to compute the \emph{Dynamic} set, and implement the computation of the \emph{Heuristic} set in Python.
Next, our implementation splits the input program into two Wasm binaries (Section~\ref{s:split}), which we implement in Rust.
In addition to two newly created Wasm binaries, our implementation also creates JavaScript glue code to dynamically link the two Wasm binaries.
To resolve the circular dependency between the two binaries, the JavaScript code creates a closure that calls the target function, and instantiates the replay binary with this closure as an import.
The record and replay step uses the existing Wasm-R3~\cite{wasm-r3} (commit hash 79b310b).
Finally, our implementation combines the target function with the replay functions (Section~\ref{s:combine}) using the wasm-merge tool~\cite{wasm-merge}.
To speed up the reduction process, our implementation tries to reduce the input program for different target functions in parallel.
We ran all experiments on an Ubuntu 24.04 LTS system with an Intel Core i9-13900K (32 logical cores) and 192 GB of DRAM.

\subsection{RQ1. Effectiveness}
\label{s:rq1}

We evaluate the effectiveness of our approach by applying \tech{} and \hybrid{}, as well as the two baselines, to the Wasm programs in Table~\ref{t:eval_set}.
We run each technique either until it terminates or a timeout of 24 hours is reached.
We run eight reduction tasks in parallel, allowing each task to use a maximum of four logical cores.
This ensures a fair comparison between wasm-reduce, \tech{}, and \hybrid{}, which all exploit parallelism.
\footnote{Wasm-shrink does not exploit parallelism.}

\begin{table*}[t]
  \centering
  \small
  \setlength{\tabcolsep}{8pt}
  \caption{\label{t:rq1}Comparison of wasm-shrink, wasm-reduce, \tech, and \hybrid.
  The most reduced programs are marked in bold.
  ``Average'' means geometric mean for effectiveness and arithmetic mean for efficiency.
  For \tech{}, ``All'' measures all functions in the reduced program, while ``Target'' measures only the target function in the reduced program.}
  \begin{tabular}{lrrrrr|rrrrc}
        \toprule
        \multirow{2}{*}{Name} & \multicolumn{5}{c}{{Effectiveness: Reduced code size (lower is better)}} & \multicolumn{4}{c}{{Efficiency: Time taken (s) (lower is better)}} \\
        \cmidrule(lr){2-6} \cmidrule(lr){7-10}
        \multicolumn{1}{c}{{}} & \multicolumn{2}{c}{{Baselines}} & \multicolumn{3}{c}{{Our Work}} & \multicolumn{2}{c}{{Baselines}} & \multicolumn{2}{c}{{Our Work}} \\
        \cmidrule(lr){2-3} \cmidrule(lr){4-6} \cmidrule(lr){7-8} \cmidrule(lr){9-10}
        \multirow{2}{*}{} & \multirow{2}{*}{\makecell{Wasm-\\shrink}} & \multirow{2}{*}{\makecell{Wasm-\\Reduce}} & \multicolumn{2}{c}{RR-Reduce} & \multirow{2}{*}{\makecell{Hybrid-\\Reduce}} & \multirow{2}{*}{\makecell{Wasm-\\shrink}} & \multirow{2}{*}{\makecell{Wasm-\\Reduce}} & \multirow{2}{*}{\makecell{RR-\\Reduce}} & \multirow{2}{*}{\makecell{Hybrid-\\Reduce}} \\
        \cmidrule(lr){4-5}
        \multicolumn{1}{c}{} & \multicolumn{1}{c}{} & \multicolumn{1}{c}{} & \multicolumn{1}{c}{All} & \multicolumn{1}{c}{Target} & \multicolumn{1}{c}{} & \multicolumn{1}{c}{} & \multicolumn{1}{c}{} & \multicolumn{1}{c}{} & \multicolumn{1}{c}{} \\
        \midrule
        wasmedge\#3018   & 4.65\%          & 1.25\%                     & 23.73\%  & 17.09\%           & \textbf{0.58\%}            &       128 &        23 &        25 &         51 \\
        wamr\#2789       & \textbf{0.05\%} & \textbf{0.05\%}            & 2.33\%   & 0.59\%            & \textbf{0.05\%}            &        24 &         6 &     1,200 &      1,210 \\
        wasmedge\#3019   & 7.71\%          & \textbf{0.06\%}            & 3.75\%   & 0.92\%            & \textbf{0.06\%}            &        10 &     4,194 &        30 &         56 \\
        wamr\#2862       & 2.95\%          & \textbf{0.18\%}            & 9.27\%   & 7.12\%            & \textbf{0.18\%}            &       299 &        54 &       136 &        187 \\
        wamr\#2450       & 11.48\%         & \textbf{0.03\%}            & 3.33\%   & 1.73\%            & \textbf{0.03\%}            &        34 &        31 &        45 &         58 \\
        wasmedge\#3076   & 27.33\%         & 0.04\%                     & 2.81\%   & 0.33\%            & \textbf{0.04\%}            &    86,400 &     1,017 &       783 &        799 \\
        mandelbrot      & 27.00\%         & 21.43\%                    & 94.70\%  & 2.57\%            & \textbf{0.43\%}            &    15,178 &    86,400 &       235 &     73,482 \\
        pathfinding     & 10.57\%         & 0.04\%                     & 31.18\%  & 0.12\%            & \textbf{0.03\%}            &     3,721 &     3,627 &       964 &      1,155 \\
        pacalc          & 17.84\%         & 0.22\%                     & 14.57\%  & 0.89\%            & \textbf{0.08\%}            &       615 &     1,014 &        15 &     21,424 \\
        wasmedge\#3057   & 26.48\%         & \textless{}0.01\%          & 2.52\%   & 0.17\%            & \textbf{\textless{}0.01\%} &    86,400 &     2,598 &     2,034 &      1,964 \\
        guiicons        & 42.38\%         & \textbf{11.65\%}           & 60.40\%  & 42.01\%           & 11.84\%                    &     1,284 &    29,266 &        75 &     20,458 \\
        rtexviewer      & 1.36\%          & 0.17\%                     & 3.44\%   & 2.28\%            & \textbf{0.02\%}            &       774 &       196 &       485 &        640 \\
        rfxgen          & 30.59\%         & 8.96\%                     & 54.17\%  & 30.21\%           & \textbf{8.72\%}            &    24,208 &     9,475 &       108 &     13,283 \\
        riconpacker     & 35.89\%         & \textbf{8.23\%}            & 42.65\%  & 35.76\%           & 8.39\%                     &     1,506 &    68,186 &        10 &     20,522 \\
        rguistyler      & 36.34\%         & 8.16\%                     & 68.11\%  & 35.20\%           & \textbf{8.15\%}            &     2,126 &    34,934 &        65 &      9,681 \\
        rguilayout      & 38.36\%         & \textbf{8.10\%}            & 62.92\%  & 37.92\%           & 10.52\%                    &    37,380 &    34,244 &        82 &     10,538 \\
        jqkungfu        & 3.13\%          & 2.78\%                     & 5.25\%   & 4.56\%            & \textbf{0.26\%}            &    86,400 &     1,241 &         9 &        183 \\
        bullet          & 62.64\%         & 6.76\%                     & 2.75\%   & 0.06\%            & \textbf{\textless{}0.01\%} &    86,400 &    86,400 &    13,585 &     17,276 \\
        funky-kart      & 36.11\%         & 5.76\%                     & 19.65\%  & 17.59\%           & \textbf{5.44\%}            &    86,400 &    86,400 &       329 &     86,400 \\
        sqlgui          & 21.66\%         & 12.23\%                    & 13.99\%  & \textbf{5.39\%}   & 11.16\%                    &    12,192 &    13,534 &        24 &     50,425 \\
        hydro           & 0.88\%          & 0.48\%                     & 2.59\%   & 1.44\%            & \textbf{0.13\%}            &    86,400 &       541 &     2,237 &      2,042 \\
        figma-startpage & 17.86\%         & \textbf{\textless{}0.01\%} & 3.24\%   & 0.02\%            & \textbf{\textless{}0.01\%} &    86,400 &       123 &        12 &         25 \\
        sandspiel       & 96.00\%         & \textbf{\textless{}0.01\%} & 0.13\%   & 0.03\%            & \textless{}0.01\%          &     2,570 &    84,626 &       897 &        952 \\
        parquet         & 25.44\%         & \textbf{\textless{}0.01\%} & 0.94\%   & \textless{}0.01\% & \textbf{\textless{}0.01\%} &    86,400 &       216 &        18 &         25 \\
        commanderkeen   & 39.51\%         & 33.44\%                    & 1.35\%   & 0.02\%            & \textbf{\textless{}0.01\%} &    86,400 &    86,400 &       788 &      5,945 \\
        jsc             & 64.51\%         & 6.36\%                     & 6.34\%   & 3.94\%            & \textbf{0.79\%}            &    86,400 &    86,400 &        65 &     15,633 \\
        boa             & 8.90\%          & 7.21\%                     & 1.36\%   & 0.94\%            & \textbf{0.64\%}            &    28,437 &    86,400 &       110 &      3,140 \\
        ffmpeg          & 4.47\%          & \textless{}0.01\%          & 0.62\%   & 0.11\%            & \textbf{\textless{}0.01\%} &    86,400 &       982 &        48 &        150 \\
        \midrule
        Average         & 12.88\%         & 0.43\%                     & 6.67\%   & 1.20\%            & \textbf{0.13\%}            &    38,603 &    28,876 &       871 &     12,775 \\
       \hline
       \end{tabular}
\end{table*}

The left-hand side of Table~\ref{t:rq1} shows the results.
On average, \tech{} reduces programs to \rrreducecode\% of their original code size (using the ``All'' size) and to \rrreducetarget\% when considering the ``Target'' size.
For comparison, the baseline reduction tools reduce programs to \wasmshrinkcode\% of their original size (wasm-shrink) and \wasmreducecode\% (wasm-reduce).
That is, \tech{} alone clearly outperforms one of the two baselines, and is competitive with the other.

The real strength of \tech{} is revealed when it is combined with existing program reduction techniques into a hybrid approach.
\hybrid{} reduces the input program to \hybridcode\% of its original size, which is a \hybridimprovesize$\times$ improvement over the current state of the art, wasm-reduce.
\hybrid{} achieves this by giving the best result in \rrhybridbest~cases, of which \rrhybridtie~are ties with the current state of the art.
The benefit of the \hybrid{} is not only in its better average performance, but also in its effectiveness on programs where the current state of the art struggles.
Table~\ref{t:highlight} highlights the difference.
Specifically, wasm-shrink produces programs larger than 100KB in 15 cases, and wasm-reduce does so in three cases.
In contrast, all \hybrid{}-reduced programs are well below 100KB, and often even further.

\begin{table}[ht]
  \centering
  \caption{Programs for which the current state of the art struggles.
  Reduced programs larger than 100 KB are shown in red, and those below 1 KB in blue.}
  \label{t:highlight}
  \normalsize
  \begin{tabular}{lccc}
    \toprule
    & Input & Wasm-reduce & Hybrid-Reduce \\
    \midrule
    mandelbrot    & 64\,KB  & 13\,KB    & \textcolor{blue}{276\,B}    \\
    bullet        & 536\,KB & 36\,KB    & \textcolor{blue}{53\,B}     \\
    hydro         & 720\,KB & 3.4\,KB   & \textcolor{blue}{939\,B}    \\
    commanderkeen & 3.9\,MB & \textcolor{red}{1.3\,MB} & \textcolor{blue}{158\,B}    \\
    jsc           & 4.3\,MB & \textcolor{red}{276\,KB} & 34\,KB                     \\
    boa           & 5.2\,MB & \textcolor{red}{374\,KB} & 33\,KB                     \\
    \bottomrule
  \end{tabular}
\end{table}

\begin{tcolorbox}
\tech{} reduces programs to \rrreducetarget{}\% of their original size.
\hybrid{} further reduces programs down to \hybridcode{}\%, which means it outperforms the state of the art by \hybridimprovesize{}$\times$ in terms of effectiveness.
\end{tcolorbox}

\subsection{RQ2. Efficiency}
\label{s:rq2}

\begin{figure*}
        \includegraphics[width=\linewidth]{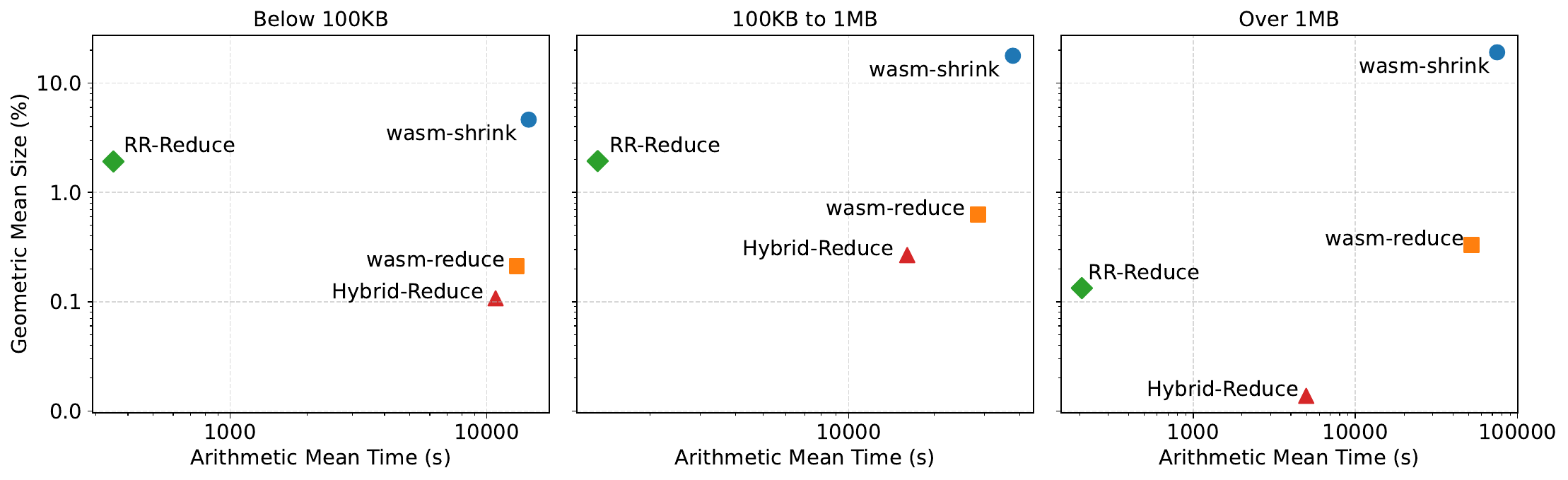}
        \caption{Tradeoff between time taken to reduce programs and the size of the reduced programs.}
        \label{f:tradeoff}
\end{figure*}

Together with effectiveness, efficiency is another important goal of program reduction techniques.
This is because, if program reduction techniques take unreasonable time to reduce the program, they are not helpful to engine developers.
The right-hand side of Table~\ref{t:rq1} shows the efficiency results, which indicate the time taken to reduce the bug-inducing programs.
\tech{} takes 14.5 minutes to reduce the bug-inducing programs, using the arithmetic mean.
\hybrid{} takes about 3.5 hours, while the current state of the art, wasm-reduce, takes eight hours.
Thus, \hybrid{} achieves a \hybridimprovetime$\times$ improvement in terms of time over the current state of the art, while also being more effective (see RQ1).
In comparison to wasm-shrink, which takes ten hours, \hybrid{} achieves a \hybridimprovetimeshrink$\times$ improvement.

\tech{} and \hybrid{} are able to achieve this speedup due to their heuristic identification of the target function.
Among the \evalsize{} programs in the evaluation set, the target function is identified in the \emph{Heuristic} set in 13 cases, the \emph{Dynamic} set in 6 cases, and the \emph{All} set in 9 cases.
As the average size of the \emph{Heuristic} set is 2.68, the \emph{Dynamic} set is 277.04, and the \emph{All} set is 1,995.57, identifying the target function in the \emph{Heuristic} set greatly reduces the search space.
Even if the heuristic fails, \tech{} can fall back to the \emph{Dynamic} or \emph{All} sets, losing some efficiency but without affecting its effectiveness.

To clearly show the tradeoff between time and reduction rate, Figure~\ref{f:tradeoff} plots time on the x-axis and size of the reduced code (``Target'' size in the case of \tech{}) on the y-axis.
We divide the evaluation set into three groups based on input code size: the below 100KB group, which contains seven programs; the 100KB to 1MB group, which contains 16 programs; and the over 1MB group, which contains five programs.
In all three groups, \hybrid{} outperforms wasm-reduce and wasm-shrink in both time and reduction rate.
\tech{} is the fastest in all three groups, while outperforming wasm-shrink in terms of effectiveness in all three groups and outperforming wasm-reduce in terms of effectiveness in the over 1MB group.
We envision both of our approaches being useful depending on the usage scenario:
For fast isolation of the bug-triggering function within minutes, \tech{} is the way to go; if a user is willing to wait a few hours, then \hybrid{} is preferable.

\begin{tcolorbox}
  \tech{} and \hybrid{} reduce programs in 14.5 minutes and 3.5 hours, respectively, which corresponds to an \rrreduceimprovetime{}$\times$ and \hybridimprovetime{}$\times$ improvement over the state of the art in terms of efficiency.
\end{tcolorbox}

\subsection{RQ3. Qualitative Analysis}
\label{s:rq3}


Our final research question qualitatively analyzes the differences between our approaches and the existing program reduction techniques.
Large programs are where the engine developers benefit the most from the help of program reduction techniques, because for them, developers need to spend the most time.
We hence focus on the four programs that are over 1MB in size.
They reveal four different bugs, providing a demonstration of the different approaches employed by program reduction techniques.
We exclude the analysis of jsc, as it triggers the same bug as boa, and much of the analysis would be redundant.

Boa, which is 5.2MB in size, triggers a wrong-code bug\footnote{Fixed by https://github.com/titzer/wizard-engine/commit/6d2b057} that gets triggered when the interpreter tries to interpret a jump with an offset bigger than 32,768 bytes.
Wasm-reduce reduces the input program to 317KB, which is still too large to manually debug.
It also contains 87 functions besides the function where the jump happens, which further complicates debugging.
Wasm-reduce cannot simply remove these functions because other functions are necessary for the jump to happen.
In contrast, \tech{} takes a different approach:
It deletes the bodies of 453 executed functions and replaces the bodies of eight functions with replay functions.
Among the nine functions that are executed in the reduced program, seven are short functions that update the state, and one is the entry to the reduced program, which calls the bug-triggering functions.
The remaining one is the bug-triggering function, which is identical to the function in the input program.
The biggest difference between the existing approaches and \tech{} is most evident in its treatment of the functions that interact with the target function to create the necessary state to trigger the bug.
In the existing approaches, every code that has been executed to create such a state is preserved.
However, \tech{} just keeps the effects of the other code to make the jump happen, which yields a program with 48,862 bytes in terms of the target size.
\hybrid{} reduces the program even further to 33,358 bytes.

Commanderkeen, which is 3.9MB in size, triggers a subtle wrong-code bug\footnote{Fixed by https://github.com/titzer/wizard-engine/commit/bc135ad} in the JIT compiler.
Wasm-reduce struggles the most, only reducing the input program to \commanderkeensize{}MB, which is hard for a human to manually debug.
The bug is a subtle register allocation problem that occurs within the engine's implementation of the \code{call\_indirect} instruction.
\tech{} selects the function where the behavior diverges as a target function and replaces the bodies of the functions that interact with the target function with replay functions, while deleting all the other functions, resulting in a reduced program of 782 bytes in terms of the target size.
\hybrid{} reduces the further down to 158 bytes, shown in Figure~\ref{f:commanderkeen-hybrid}.

Ffmpeg, which is 5.3MB in size, triggers a compiler-crash bug\footnote{Fixed by https://github.com/titzer/wizard-engine/commit/4e3e221}.
As this is a compiler-crash bug, most of the functions can be safely deleted except the bug-triggering function.
Thus, wasm-reduce is effective in this case, reducing the input program to 479 bytes in 982 seconds.
\hybrid{} achieves an even smaller result of 45 bytes in only 150 seconds.

Parquet, which is 1.7MB in size, triggers a trivial compiler-crash bug\footnote{Fixed by https://github.com/titzer/wizard-engine/commit/33ec201} in the Wizard engine.
For this bug to be triggered, it is sufficient to contain a single memory declaration with 64-bit addresses.
All other code can be removed as long as the Wasm module retains the offending memory declaration.
Thus, wasm-reduce is effective, producing a reduced program that contains a single function with a single nop instruction in 216 seconds.
\hybrid{} achieves the same reduction but quicker; it takes only 25 seconds to obtain the reduced output.

\begin{tcolorbox}
  For two programs over 1MB causing wrong-code bugs, only \hybrid{} is effective enough to facilitate the manual debugging while still being efficient.
  For two programs over 1MB causing compiler-crash bugs, the current state-of-the art, wasm-reduce, is effective, but \hybrid{} gives even better results faster.
\end{tcolorbox}

\section{Discussion}
\label{sec:discussion}

\subsection{Threats To Validity}
\label{sec:threats}

\paragraph{Internal validity}
Our effectiveness results are influenced by the timeout setting we used (24 hours).
Thus, effectiveness would improve if we allow the tools to run longer.
However, we believe 24 hours is a reasonable assumption on what developers will tolerate when debugging a single bug-triggering program. This timeout is also used in other recent program reduction work~\cite{DBLP:conf/sigsoft/ZhangX00S23}.

\paragraph{External validity}
Our results are limited to the 28 programs in our evaluation set and may not generalize to all Wasm applications.
However, our evaluation set covers diverse real-world use cases, including programming-language runtimes, media applications, video games~\cite{wasm-r3}, and automatically generated Wasm programs~\cite{10.1145/3650212.3680358}.

\subsection{Limitations}
\label{sec:limitations}

\tech{} uses simple string search over the engine’s output to heuristically identify a target function that triggers the bug.
This heuristic can be effective for engine crash bugs and wrong-code bugs that lead to a runtime error.
It is not effective for wrong-code bugs that do not manifest as observable changes in behavior.
In such cases, \tech{} can fall back to the Dynamic set and All set at the cost of longer reduction time, yet still produce the same final minimized output.
Another limitation is that \tech{} currently supports reduction of Wasm programs up to the Wasm 2.0 specification, excluding SIMD, due to the limitations of Wasm-R3.
It also relies on the correctness and performance of Wasm-R3.
Addressing more Wasm extensions and improving performance will require further engineering of the Wasm-R3 toolchain.

\subsection{Generalization to Other Programming Languages}

Conceptually, \tech{} can be applied to other programming languages.
There are a few conditions for this to be possible:
(1) A record and replay implementation must exist.
(2) The record and replay system must support selective record and replay.
(3) The output of selective record and replay must be a regular program.
\section{Related Work}
\label{sec:related}

\paragraph{Dynamic Analysis for WebAssembly}

There are several dynamic analysis techniques for Wasm.
Wasabi~\cite{Wasabi} designs and implements bytecode-level dynamic instrumentation for Wasm to enable diverse dynamic analyses.
Wizard~\cite{titzer2024flexible} supports non-intrusive instrumentation for Wasm by engine-level dynamic instrumentation.
Wasm-R3~\cite{wasm-r3} is a record and replay technique for Wasm that enables the generation of executable, standalone Wasm benchmarks.
Wemby~\cite{Draissi2025WembysWH} utilizes dynamic analysis to detect memory corruption bugs in Wasm.
Our work contributes to the field by utilizing the execution behavior of Wasm programs to improve Wasm program reduction.

\paragraph{Record and Replay}

Record and replay is a well-established research area that has been studied in multiple domains, including architectural support~\cite{FDR}, OS-level implementations~\cite{revirt}, user-space implementations~\cite{rr}, language-runtime integrations~\cite{RANDR}, JavaScript benchmark generation~\cite{jsbench}, and Wasm benchmark generation~\cite{wasm-r3}.
Selective record and replay techniques~\cite{Orso2005SelectiveCA, DBLP:conf/kbse/SaffAPE05, jalangi, DBLP:conf/issta/BurgerZ11, Hammoudi2015OnTU}, which record and replay only part of an execution, are particularly relevant to us.
All of these techniques utilize selective record and replay with different goals in mind.
To our knowledge, however, \tech{} is the first to apply selective record and replay to program reduction.

\paragraph{Test Input Reduction}

Delta debugging (DD)~\cite{DBLP:journals/tse/ZellerH02} pioneered automated test input reduction.
Hierarchical Delta Debugging (HDD)~\cite{DBLP:conf/icse/Misherghi06} adapts DD for hierarchical inputs such as programming languages.
Generalized Tree Reduction (GTR)~\cite{Herfert2017AutomaticallyRT} generalizes HDD to support arbitrary tree transformations and specializes these transformations by learning from a corpus of example data.
Perses~\cite{Sun2018PersesSP} uses formal syntax of the program to make language-agnostic program reduction more effective, efficient, and general.
Vulcan~\cite{DBLP:journals/pacmpl/XuTZZJS23} utilizes diverse program transformations to further reduce the output of language-agnostic program reducers.
PPR~\cite{DBLP:conf/sigsoft/ZhangX00S23} minimizes pairs of programs rather than a single program.
LPR~\cite{DBLP:conf/issta/Zhang0X0TS24} leverages LLM for program reduction.
None of these approaches, however, exploits the execution behavior of the input program to guide reduction.
In contrast, our approach is \trait{}; it leverages an execution of the program to achieve more effective and efficient outcomes.

There is one approach that we are aware of that exploits execution behavior to improve delta debugging.
Fast-Reduce, one of the program reduction techniques introduced in~\cite{Regehr2012TestcaseRF}, uses run-time information obtained from instrumentation.
One transformation that Fast-Reduce applies is to inline calls to functions with their dynamic effect, which allows removal of the called function after all its call sites are inlined.
Our approach differs from Fast-Reduce in the following ways:
(1) Fast-Reduce relies on dynamic analysis specifically built to integrate with the test case generator CSmith~\cite{Yang2011FindingAU}, whereas \tech{} employs a general-purpose record and replay tool;
(2) Fast-Reduce selects an arbitrary final effect of the function to inline, whereas \tech{} records and replays every effect of the function across its multiple calls.
(3) Fast-Reduce does not empirically outperform other techniques in terms of effectiveness, whereas the use of \tech{} achieves such outcomes.
In summary, although Fast-Reduce and \tech{} share some similarities, we find \tech{} to be a more principled and effective realization of the idea of using execution behavior to improve program reduction.

\section{Conclusion}
\label{sec:conclusion}

We present an \trait{} program reduction technique for Wasm that utilizes record and replay for the purpose of program reduction, with its concrete implementation in \tech{} and \hybrid{}.
A key insight is that record and replay of only part of the input program preserves the execution behavior of that part while either deleting or replacing the rest of the program.
We evaluate \tech{} and \hybrid{} on a set of Wasm programs and found it to be effective and efficient.
We hope that \tech{} and \hybrid{} pave the way for program reduction techniques to scale to much larger and more complex programs, extending the reach of occasions where Wasm engine developers can benefit from program reduction techniques.
In addition, we hope that \tech{} and \hybrid{} inspire the development of similar techniques in other languages by leveraging selective record and replay.

\section*{Data Availability}
The artifact (source code, benchmark, evaluation outputs, and documentation) is available at https://github.com/sola-st/rr-reduce.

\section*{Acknowledgements}
We thank Laurence Tratt and anonymous reviewers for their helpful feedback.
This work was partially supported by NSF award \#2148301, the WebAssembly Research Center, the National Research Foundation of Korea (NRF) (2022R1A2C2003660 and 2021R1A5A1021944), an Institute of Information \& Communications Technology Planning \& Evaluation (IITP) grant funded by the Korea government(MSIT) (2024-00337703), Samsung Electronics Co., Ltd., by the European Research Council (ERC, grant agreements 851895 and 101155832), and by the German Research Foundation within the DeMoCo project.

\bibliographystyle{IEEEtran}
\bibliography{references,referencesMP}
\end{document}